\documentclass[natbib=true, sigconf,screen]{acmart}

\usepackage{subcaption}
\usepackage[font=small, labelfont=bf, skip=5pt]{caption}

\AtBeginDocument{%
  \providecommand\BibTeX{{%
    \normalfont B\kern-0.5em{\scshape i\kern-0.25em b}\kern-0.8em\TeX}}}

\setcopyright{acmcopyright}
\copyrightyear{2024} 
\acmYear{2024} 
\setcopyright{rightsretained} 
\acmConference[UbiComp Companion '24]{Companion of the 2024 ACM International Joint Conference on Pervasive and Ubiquitous Computing }{October 5--9, 2024}{Melbourne, VIC, Australia}
\acmBooktitle{Companion of the 2024 ACM International Joint Conference on Pervasive and Ubiquitous Computing (UbiComp Companion '24), October 5--9, 2024, Melbourne, VIC, Australia}
\acmDOI{10.1145/3675094.3677598}
\acmISBN{979-8-4007-1058-2/24/10}

\begin{document}

\title{Inside Out or Not:  Privacy Implications of Emotional Disclosure}

\author{Elham Naghizade}
\email{e.naghizade@rmit.edu.au}
\affiliation{
  \institution{RMIT University}
  \city{Melbourne}
    \country{Australia}
}

\author{Kaixin Ji}
\email{kaixin.ji@student.rmit.edu.au}
\affiliation{%
  \institution{RMIT University}
  \city{Melbourne}
  \country{Australia}
}

\author{Benjamin Tag}
\email{benjamin.tag@monash.edu}
\affiliation{%
  \institution{Monash University}
  \city{Melbourne}
    \country{Australia}
}

\author{Flora Salim}
\email{flora.salim@unsw.edu.au}
\affiliation{%
  \institution{The University of New South Wales}
  \city{Sydney}
    \country{Australia}
}
\renewcommand{\shortauthors}{Naghizade, et al.}

\begin{abstract}

Privacy is dynamic, sensitive, and contextual, much like our emotions. Previous studies have explored the interplay between privacy and context, privacy and emotion, and emotion and context. However, there remains a significant gap in understanding the interplay of these aspects simultaneously. In this paper, we present a preliminary study investigating the role of emotions in driving individuals' information sharing behaviour, particularly in relation to urban locations and social ties. We adopt a novel methodology that integrates context (location and time), emotion, and personal information sharing behaviour, providing a comprehensive analysis of how contextual emotions affect privacy. The emotions are assessed with both self-reporting and electrodermal activity (EDA). Our findings reveal that self-reported emotions influence personal information-sharing behaviour with distant social groups, while neutral emotions lead individuals to share less precise information with close social circles, a pattern is potentially detectable with wrist-worn EDA. Our study helps lay the foundation for personalised emotion-aware strategies to mitigate oversharing risks and enhance user privacy in the digital age. 

\end{abstract}



\maketitle

\section{Introduction}

Emotions and our ability to recognise them and respond to them are fundamental to our ability to navigate our lives
(including work, education, family and social interactions). Nowadays with the help of interconnected smart devices, we are not only able to constantly capture our whereabouts, and our online social interactions, but also have access to our own physiological data, such as facial expressions, vocal cues, as well as heart rate and hormone levels. 

Such rich personal data has made it possible to anticipate the real-time recognition and prediction of one's own emotion~\cite{nandi2022} and react to individuals’ emotional responses in different contexts~\cite{breitfuss2021}. This has resulted in the emergence of emotion-aware personalised services facilitating a wide range of personalised 
~\cite{can2019stress} and contextualised ~\cite{Nepal2020Job, riaz2018} recommendations.

Emotions, however, allude to intimate and private aspects of individuals' lives. In 2021, both the European Commission (through proposing the AI Act)\footnote{\url{https://eur-lex.europa.eu/legal-content/EN/TXT/?uri=CELEX\%3A52021PC0206}} and the UN Human Rights Council\footnote{\url{https://documents.un.org/doc/undoc/gen/g21/285/95/pdf/g2128595.pdf?token=ucMNChg4dVP7tonlxK&fe=true}} enlisted emotion recognition technologies as emergent priorities in the global right to privacy.
Despite this, there is still little known about the impact of emotions on our personal information sharing and the privacy of our emotions altogether. 
This paper aims to bridge this gap by focusing on the first question: \textit{If and how emotions influence individuals' personal information disclosure behaviour and explain shifts in privacy preferences in digital interactions?} To better contextualise our study, we focus on emotions in urban environments and investigate how emotions, influenced by time and location, affect an individual's willingness to share private (location) information.

Understanding the emotional impact of urban environments is an extensively studied area since it can guide the design of more liveable and supportive cities. Whilst this is a well-established field of research, recent advancements in sensor and emotion recognition technologies have significantly driven innovative explorations in this domain~\cite{pykett2020, resch2020, han2023stress}. On the contrary location privacy has been considered one of the most urgent areas to study since revealing
an individual’s location information may have negative implications, e.g., by exposing oneself to potentially being targeted
for unsolicited location-based advertisements or even being stalked and physically harmed~\cite{krumm2009}. Whilst a large number of studies in this area focuses on developing robust privacy-preserving techniques to protect users' sensitive location information from unauthorised access and inference attacks~\cite{liu2018}, several user-centric studies have highlighted the importance of transparency and user control, leading to the development of personalised nudging approaches that aim to empower users to make informed privacy decisions~\cite{rivadeneira2023, Jackson2018Notifications, fitness2022Velykoivanenko}.

In the context of emotion and privacy, social media represents a particularly dynamic and influential environment. Several studies have examined the role of emotions on information sharing in social media~\cite{robertson2023, Stieglitz2013}. A recent study in this area highlights that emotional arousal significantly increases the likelihood of sharing information online. Whilst positive emotions, such as happiness or excitement, tend to make users view the quality of information more favourably, hence increasing their tendency to share it, negative emotions, such as anger or fear, can also drive information sharing by reinforcing existing beliefs, especially with controversial or sensational content~\cite{Hasell2023}. This information, however, is not necessarily private information but rather content available online. On social media, on the other hand, emotions are mainly measured by analysing the sentiment of users' posts or content that is being shared rather than by analysing their \textit{contextual, instantaneous} emotions.

This study aims to bridge this gap by 
conducting a lab experiment using simulated contextual emotions through video recordings. It collects self-reported emotions, electrodermal activity (EDA), as well as decisions about sharing private location information with specific social groups from 20 participants.
Our findings reveal that self-reported emotion influences personal information sharing behaviour with distant social groups while neutral emotions lead individuals to share less precise
information with close social circles. This relation can possibly be detected using wrist-worn EDA.

\section{Related Work}

\paragraph{ Emotion \& Information Sharing} 
Whilst the study in~\cite{Stieglitz2013} highlights that emotionally charged content was shared more and faster on Twitter, a recent study focusing on political news sharing suggests that negative news content triggers anger and sad reactions in users and tends to be shared more. On the contrary, positive emotional reactions seem to decrease the news-sharing trend~\cite{Leon2021}. These studies, however, focus on (re)sharing emotionally charged content on social media rather than examining the impact of the user's emotions on \textit{personal} information sharing. ~\cite{Hasell2023}, in their recent study focus on the emotion of the sharers and argue that the more emotionally aroused, the more likely that individuals share information on social media, especially to seek social support, social influence or retribution. Their findings, however, do not focus on the role of sharer's \textit{momentary} emotion in their personal information-sharing behaviour. \citet{zhang2020privacy} conducted a study to investigate the impact of emotions on privacy management strategies and personal information-sharing behaviours. This survey-based study was carried out with 556 university students exploring if stress was used as an indicator of emotion. The results showed that the level of stress is negatively correlated to privacy concerns, but the motivation is varied by gender: while male participants were sharing more frequently under stress, female participants were making more intimate disclosures.

\paragraph{Location Information \& Privacy}~\citet{tang2010rethinking} propose two motivations for location sharing, namely \emph{purpose-driven} and \emph{social-driven} motivations. For purpose-driven sharing, people were more comfortable sharing their exact location because they needed to use it to receive a service, e.g., finding the closest petrol station. However, for social-driven sharing, the researchers found that users would apply different strategies to blur their location. For example, their participants prefer to share semantic location labels, such as a library or class, or share an activity, rather than fine-grained geographic information. 
To the best of our knowledge, the role of emotion on sharing personal location information is understudied. Perhaps, the closest such study is by~\citet{xie2014location}, who conducted a crowd-sourced experiment where the participants imagined themselves in some scenarios before being asked to select their privacy preferences. The privacy levels considered whether the participants would share their location with Family, Friends or Colleagues. Their experiment also included some analysis of emotion. This study found that location sharing preferences was sometimes time-sensitive, while there were only two actions found when it was influenced by emotion, sharing all or nothing. 
However, the exact relationship was beyond the scope of the project, emphasising the need to investigate this triangular relation further. 
Besides, another limitation of this survey setting might be that relying on imagination does not effectively simulate real-life situations.
Our project uses video as a stronger stimulus to create the sensation of the actual situation, and invoke emotions more reliably~\cite{Gross1995}.

\paragraph{Emotion Elicitation using Videos}
The last decade has witnessed a surge in studies that have demonstrated that videos are effective in eliciting a wide range of emotions (measured through valence and arousal dimensions) due to their ability to combine visual and auditory stimuli~\cite{Gross1995}. 
For example,~\cite{carvalho2012emotional, koelstra2011deap, Soleymani2011} have used emotionally charged film clips to invoke specific emotional states and then measured participants' responses using both subjective self-reports and objective physiological indicators. In urban scenarios,~\cite{mavros2022, osborne2017,pykett2020} have studied emotional responses to various urban stimuli in a lab setting versus in the wild. Their finding suggests that sensors can be a powerful tool for understanding individual responses to environments, but should be used cautiously.

\section{Experiment}
\label{sec:experiment}
\begin{figure}[h]
    \centering
    \includegraphics[width=\linewidth]{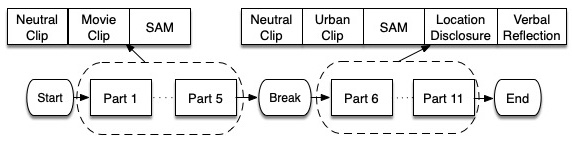}
    \caption{Experiment Procedure. The sequences on the top depict the steps taken for each part, whereas parts 1-5 and 6-11 correspond to the movie clips and urban clips. Part 1 is always the neutral clip. The rest of the videos in both sections are presented randomly. }
    \label{fig:procedure}
\end{figure}

\paragraph{\textbf{Procedure}}
The experiment takes place in a lab room that is constantly illuminated. 
The consent form is signed before the experiment starts, followed by debriefing and equipment setup, including wearing an E4 band\footnote{https://www.empatica.com/en-gb/research/e4/} to collect EDA sensor data.
The participants then need to complete a pre-study survey regarding their general privacy attitudes. We used a survey proposed in ~\cite{kehr2015b} and focused on 5 items: Global Information Privacy Concern (GIPC) and Trusting Beliefs (TB) from~\cite{malhotra2004} and Perceived Privacy Risks (PPR), Perceived Benefits of Information Disclosure (PBID) and Perceived Privacy (PP) from~\cite{dinev2013}.
As shown in Figure~\ref{fig:procedure}, the experiment consists of two sections. In both sections, the participants watch a short video, after a 15-second preparation, on a 24 inch monitor. 

After each video, the participants rate their valence and arousal using the 9-point Self-Assessment Manikin (SAM) scale~\cite{Bradley1994}.
The first section contains five movie clips, approximately 2 minutes long each, corresponding to distinct emotions. 
After the first section, a 5-minute break is given to ensure participants are not tired and minimise effects of one sensation bleeding into the next. The second section contains six 1.5-minute urban clips. Before watching the clips, participants are asked to assume they are in that location. To evaluate the participants' tendency to disclose their location information, we adopt the approach from~\cite{lin2013} and focused on two dimensions: ``What to share",  explores the precision of the location that is being shared, e.g., CBD vs. Bourke St. while ``Whom to share'', explores the preferred social groups as the recipients of the information, e.g., Close Friends and Family vs Friends on Social Media.
The participants are also encouraged to speak out if they have any further explanations for their choices.

\vspace{-5pt}

\paragraph{\textbf{Movie Clips}} We have selected a subset of five clips from~\cite{Soleymani2011} with the aim to capture distinct emotions. The original dataset consists of 20 clips rated by more than 50 participants based on a 9-point SAM. We chose this subset to include the most discriminatory clips in the four quadrants of the reported valence/arousal plane as well as a neutral video~\cite{posner2005}. 

\begin{figure}[h]
    \centering
    \begin{subfigure}[b]{0.3\linewidth}
         \centering
        \includegraphics[width=\linewidth]{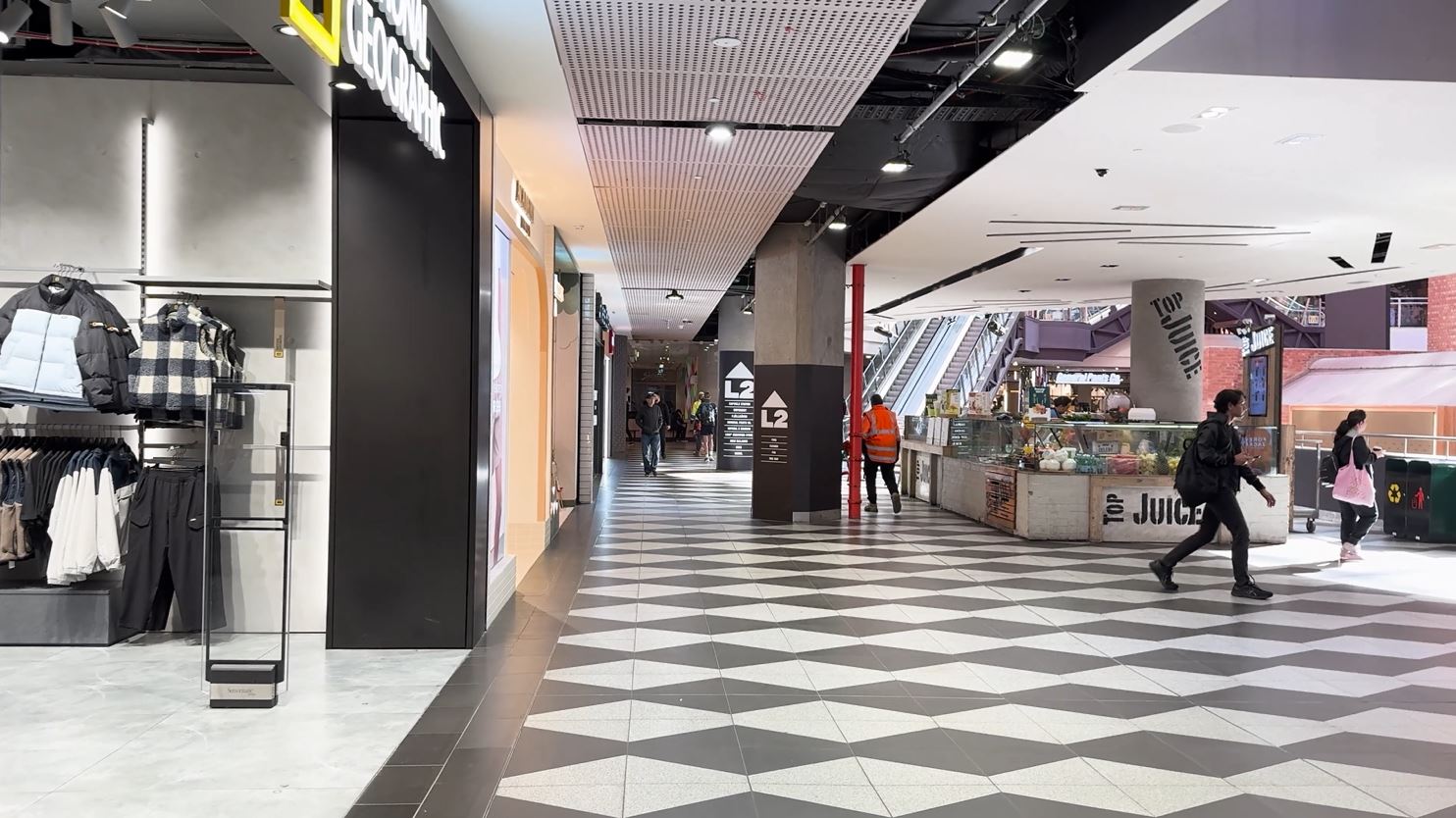}
        \caption{Indoor - Day}
     \end{subfigure}
     \begin{subfigure}[b]{0.3\linewidth}
         \centering
        \includegraphics[width=\linewidth]{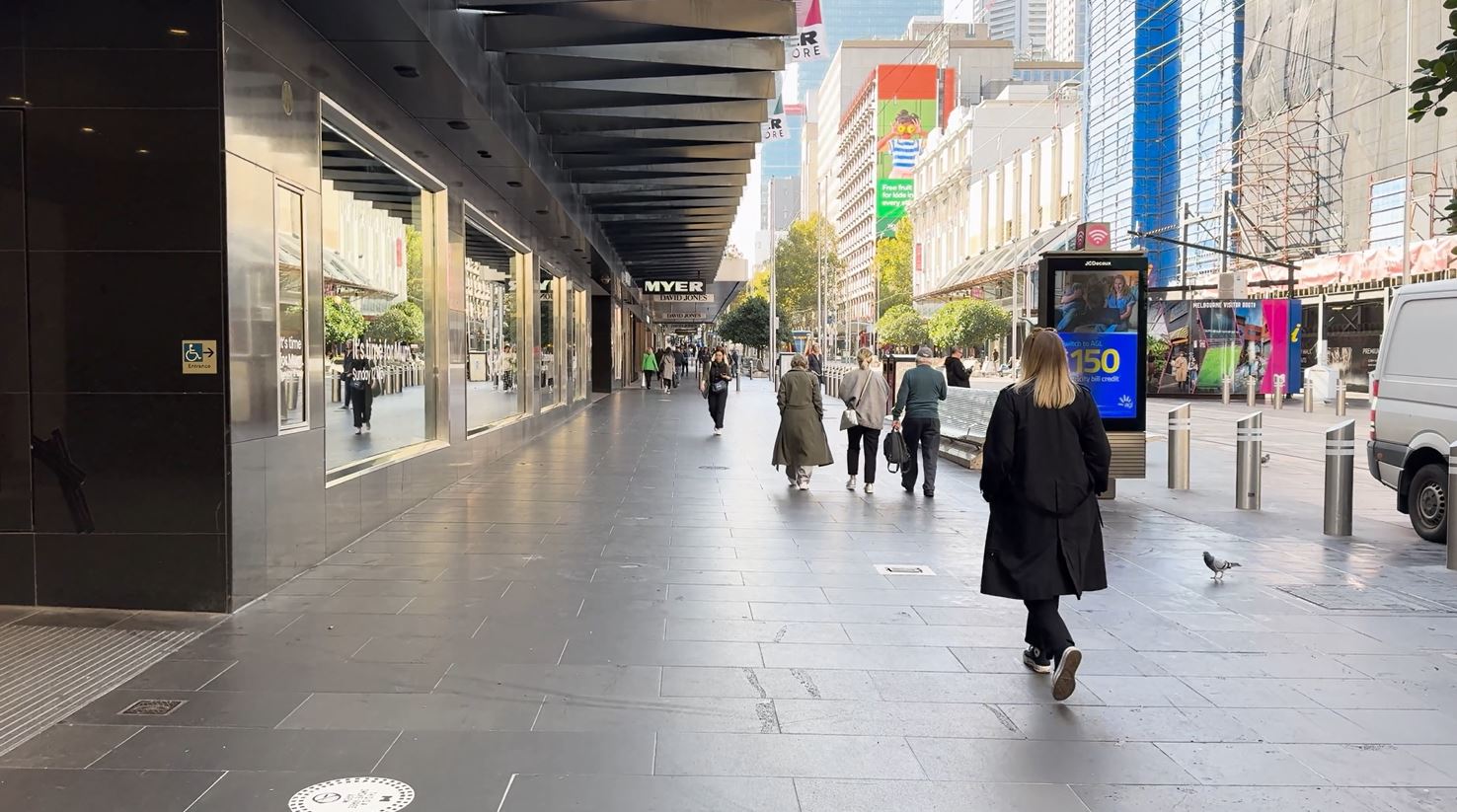}
        \caption{Outdoor - Day}
     \end{subfigure}
     \begin{subfigure}[b]{0.3\linewidth}
         \centering
        \includegraphics[width=\linewidth]{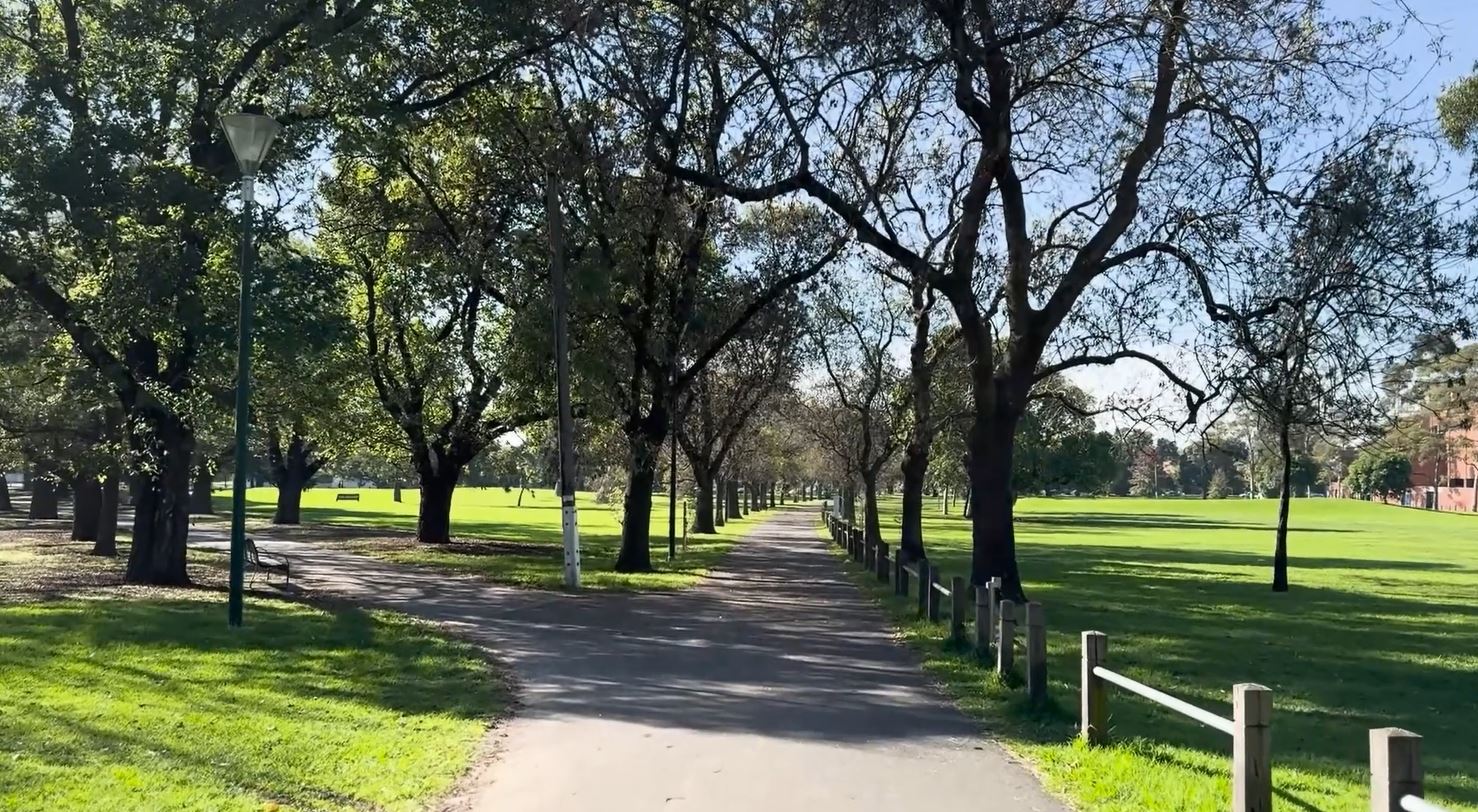}
        \caption{Greens - Day}
     \end{subfigure}
     \vfill
     \begin{subfigure}[b]{0.3\linewidth}
         \centering
        \includegraphics[width=\linewidth]{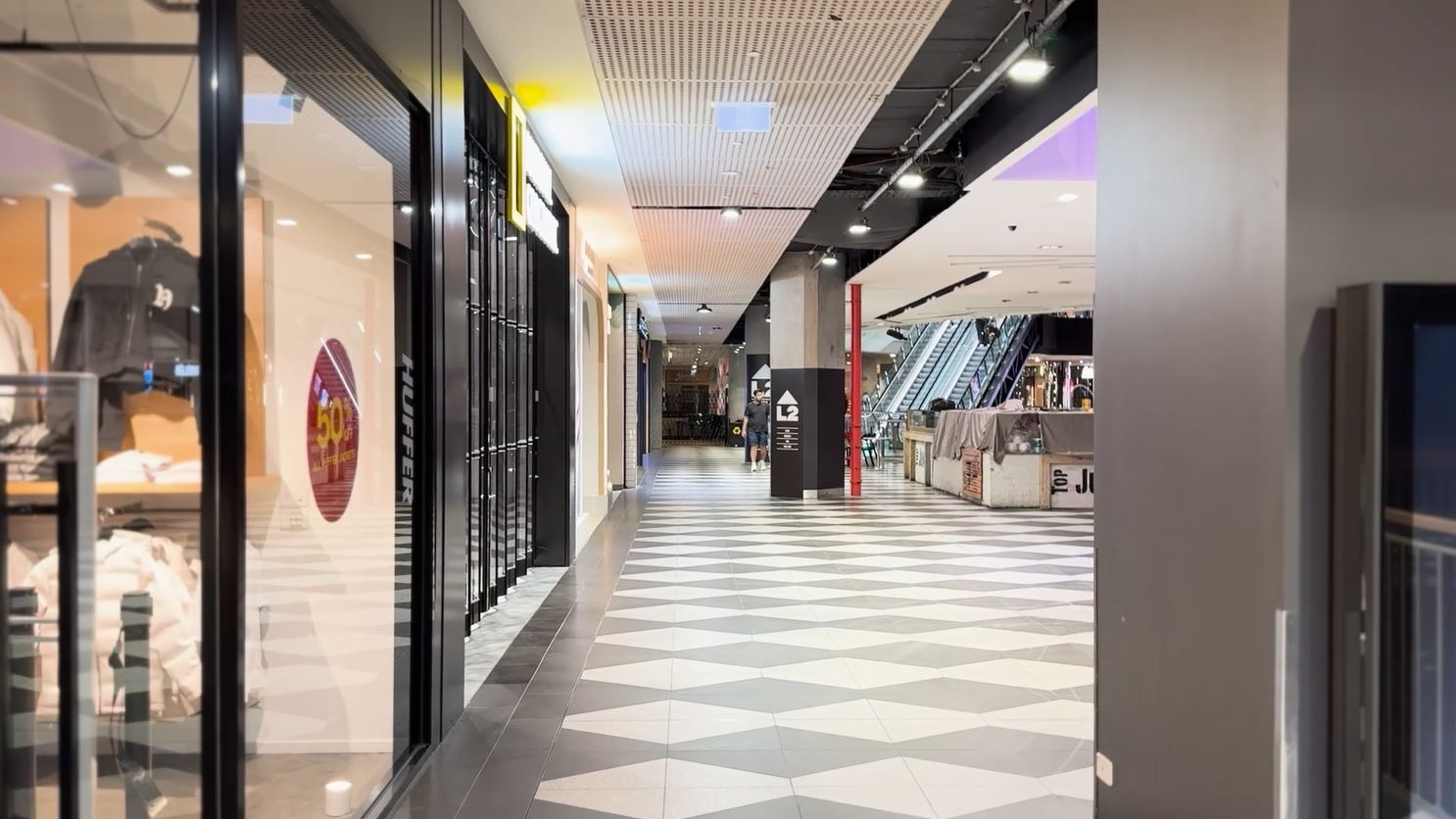}
        \caption{Indoor - Night}
     \end{subfigure}
     \begin{subfigure}[b]{0.3\linewidth}
         \centering
        \includegraphics[width=\linewidth]{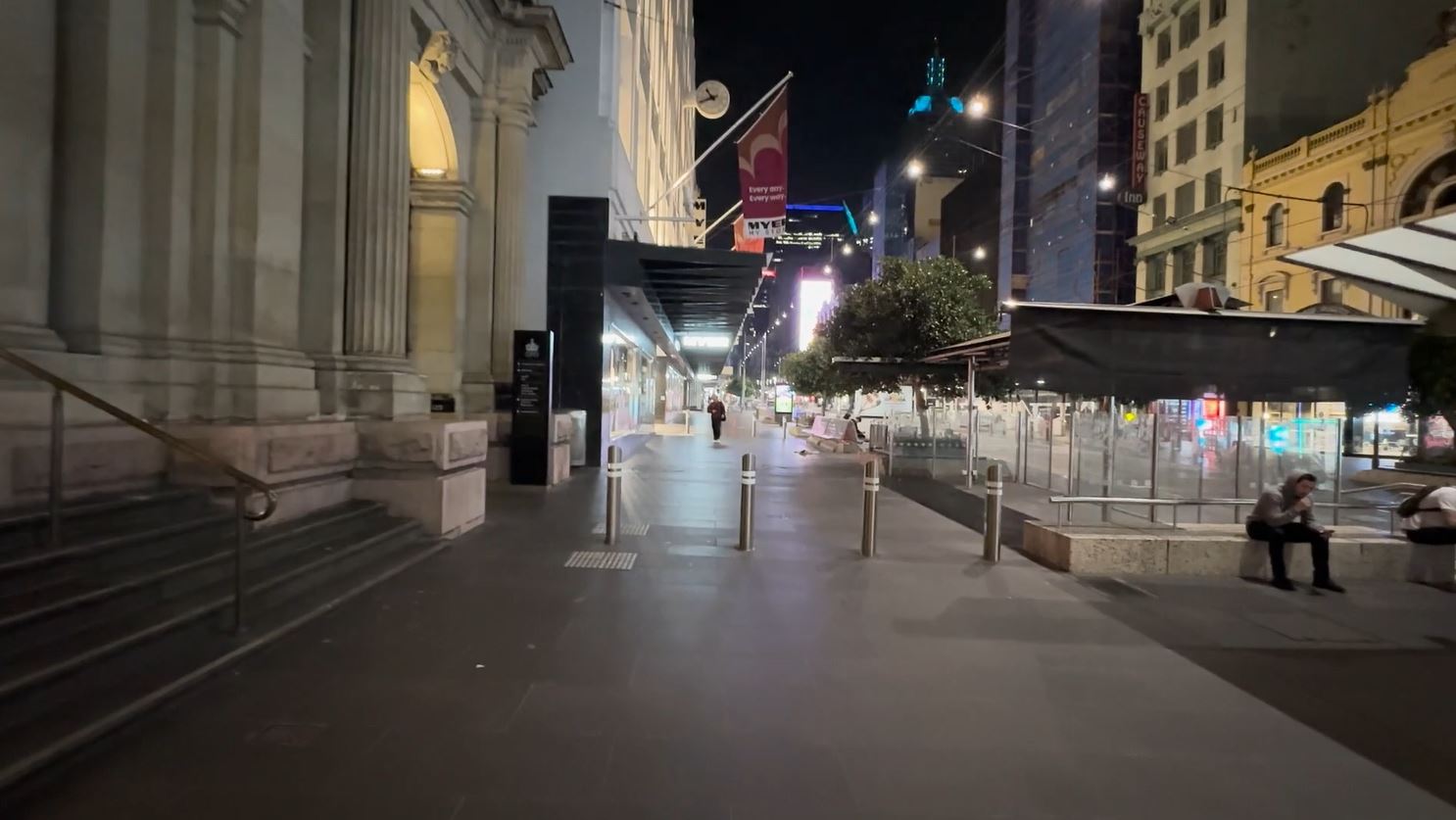}
        \caption{Outdoor - Night}
     \end{subfigure}
     \begin{subfigure}[b]{0.3\linewidth}
         \centering
        \includegraphics[width=\linewidth]{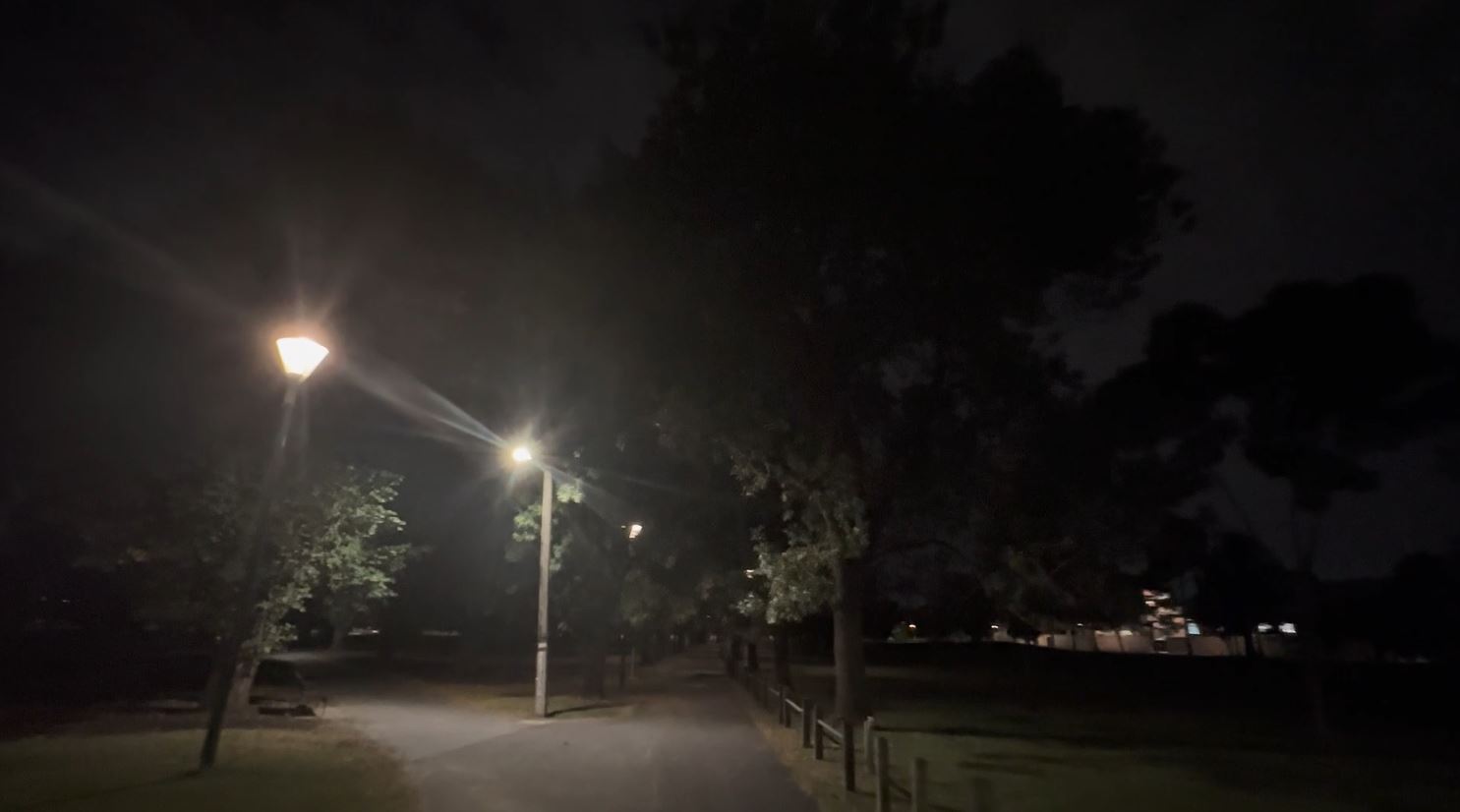}
        \caption{Greens - Night}
     \end{subfigure}
    \caption{Snapshots of the six urban clips. }
    \label{fig:screenshots}
\end{figure}

\paragraph{\textbf{Urban Clips}}
Following~\cite{mavros2022}, the research team has recorded videos in 3 different categories - urban indoor, outdoor, and green-\\spaces, during 2 specific time periods - 10-11AM and 10-11PM on weekdays. In both urban scenarios (indoor and outdoor), we have focused on landmarks that citizens can easily identify. The videos are shot from a first-person perspective using an iPhone 15 Pro.
All videos are recorded only under sunny weather. Figure~\ref{fig:screenshots} shows snapshots from these three scenarios during daytime and nighttime.

\paragraph{\textbf{Participants}}

All participants recruited have resided in Melbourne for more than six months. 
This ensures that participants are sufficiently familiar with the urban landmarks, hence minimising any impact related to the excitement or stress of being in a new environment. A total of 20 participants from the university community were recruited (10 Female, 10 Male) where 81\% and 19\% of them are between 25-34 and 18-24 years old, respectively.
\paragraph{\textbf{Pre-processing SAM}}
We first re-code the valence and arousal ratings from the SAM textual description, i.e., positive to negative and calm to agitated, to numbers ranging from 1-9 where 5 for neutral, 1 for extremely negative and 9 for extremely positive, followed by categorisation. The responses with both valence and arousal rated as 5 are labelled as `Neutral (NEU)'. The rest are divided according to $\ge5$ as `High (H)' and $\leq5$ as `Low (L)'. 

\paragraph{\textbf{Pre-processing EDA}}
The EDA signal is first cleaned and denoised with a low-pass filter and 5-second rolling median, then standardised by z-score \cite{diUnobtrusive2018}.
The signal is segmented at the beginning and the end of each clip as a trial. Baseline correction is applied by subtracting the average value of the onset second before decomposing into tonic and phasic values. Finally, each trial is averaged with a 10-second time window with 50\% overlap.

\section{Results \& Analysis}

\begin{figure}[h]
    \centering
    \includegraphics[width=0.75\linewidth]{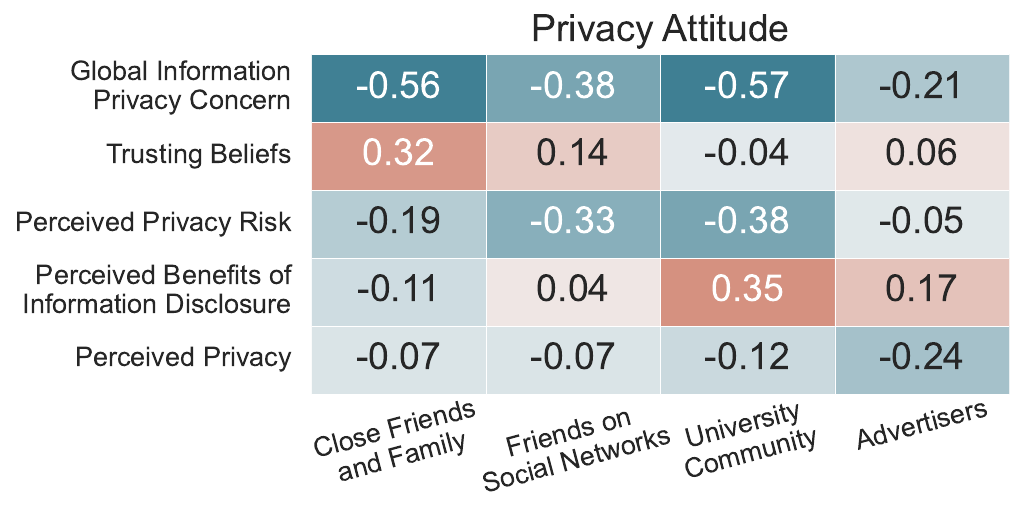}
    \caption{Pearson Correlation between privacy attitudes and the tendency of sharing location information to different recipient groups.}
    \label{fig:emotion_sharing}
\end{figure}

\paragraph{\textbf{Privacy Attitude}}
\begin{figure}[ht!]
    \centering
    \includegraphics[width=0.95\linewidth]{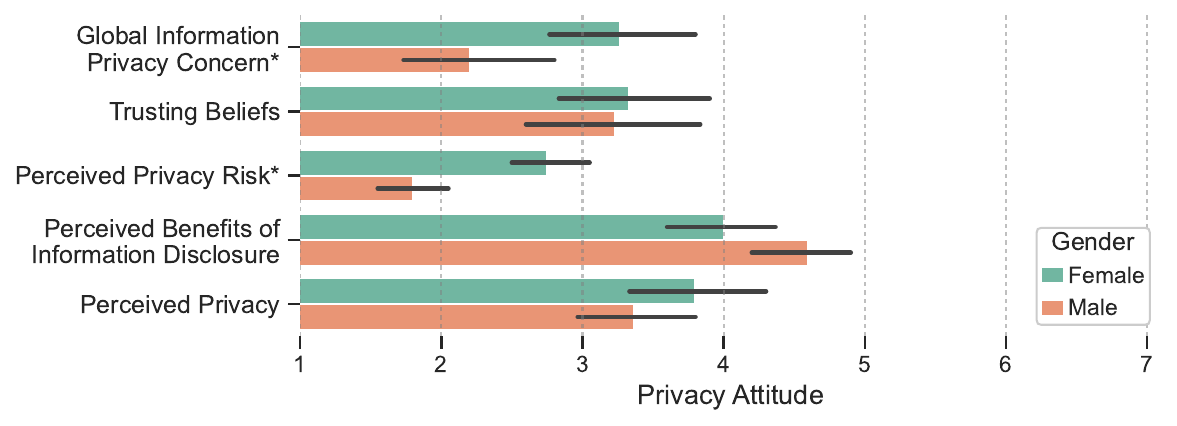}
    \caption{Self-reported privacy attitude for each of the five items. *GIPC and PPR are reversely coded.}
    \label{fig:privacy_preference}
\end{figure}
Figure~\ref{fig:privacy_preference} shows the average ratings for each item in the privacy attitude questionnaire described in Section~\ref{sec:experiment}. As can be seen, male participants are generally less concerned about their privacy and have more positive attitudes towards sharing personal information on social media while having greater doubts about the protection of their privacy on the platforms, i.e., PP.
Figure~\ref{fig:emotion_sharing} depicts the correlation between participants' average score for the five privacy attitudes items and their tendencies to share location information with different social groups (Close Friends and Family, Friends on Social Networks, University Community, Advertisers). 
GIPC and PPR are generally negatively correlated with different groups, whilst the strongest negative correlation between GIPC is observed with the University Community ($r=-0.57$) and Close Friends and Family ($r=-0.56$). TB and PBID tend to have weaker and more varied correlations, with a notable positive correlation for TB and Close Friends and Family ($r=0.32$) as well as PBID with the University Community ($r=0.35$).

\paragraph{\textbf{Impact of Urban Environment on Emotions}}

\begin{figure}[ht!]
    \centering
    \includegraphics[width=0.75\linewidth]{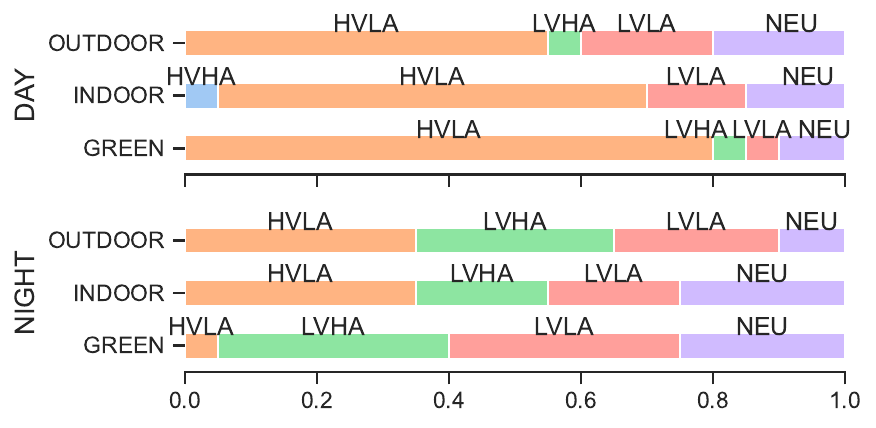}
    \caption{Self-reported SAM ratings for the six urban clips.}
    \label{fig:loc_emotion}
\end{figure}

The results of self-rated emotions are presented in Figure~\ref{fig:loc_emotion}. Overall, participants reported a majority of positive emotions in greenspace at daytime and negative emotions at nighttime.
At daytime, the participants mostly feel high valence and low arousal at all locations, with the highest valence at the greenspace ($\overline{X}: 6.7, 3.2$). 
At nighttime, they perceive the lowest valence and highest arousal at the greenspace ($\overline{X}: 4.4, 4.9$), followed by the urban outdoor ($\overline{X}: 5.1, 4.4$). 

A chi-square test of independence reveals no significant association between emotion and location, $\chi^2 (8, N = 120) = 4.16, p > .05$, but a significant difference between emotion and time, $\chi^2 (4, N = 120) = 27.30, p= .000$, with a moderate effect size, $\phi_c =.48$. 
This result is consistent when considering females and males separately, where $\chi^2 (3, N = 60) = 19.52, p= .000$ on female group, and $\chi^2 (4, N = 60) = 10.88, p= .028$ on male group. 

\paragraph{\textbf{Impact of Emotion on Location Sharing}}

\begin{figure}[ht!]
    \centering
    \includegraphics[width=0.99\linewidth]{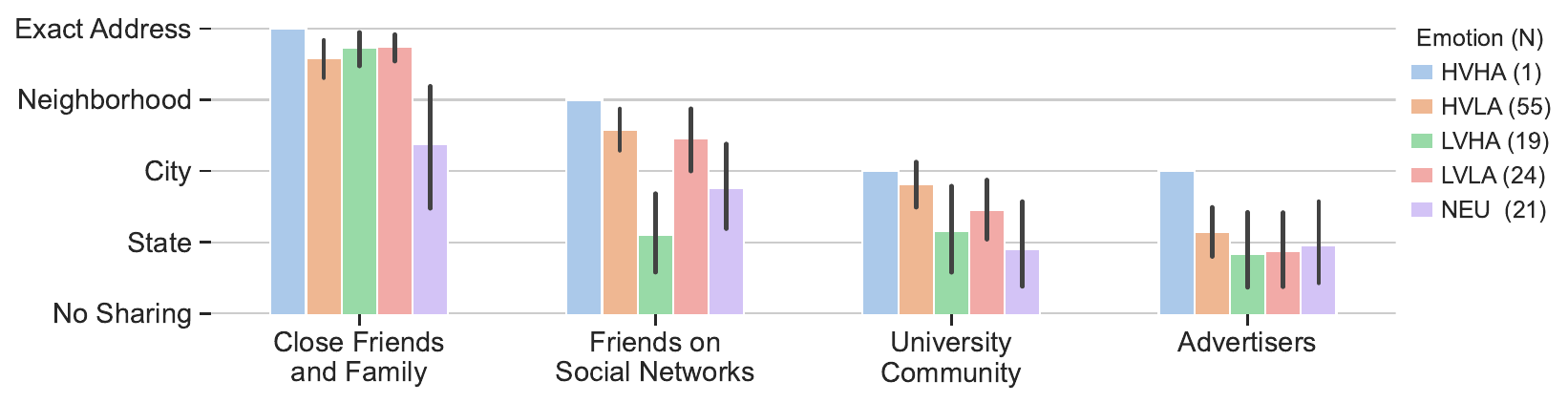}
    \caption{Distribution of emotions and level of location information shared to different social groups.}
    \label{fig:location_share}
\end{figure}

Figure~\ref{fig:location_share} shows the average number of participants sharing their location information at varying precision levels against their SAM ratings. 
Perhaps the most interesting observation is that when reporting neutral emotion, participants prefer to share less precise location information with their Close Friends and Family. The chi-square test of independence also confirms that emotions and the tendency of sharing precise location information significantly differ when sharing with Close Friends and Family, $\chi^2 (12, N = 120) = 28.32, p = .005$, with a medium effect size, $\phi_c=.28$. 

On the other hand, lower reported arousal levels (LVLA and HVLA) seem to result in sharing more precise location information with Friends on Social Media and University Community, however, the chi-square test of independence test only confirms this to be significant for the University Community ($\chi^2 (16, N = 120) = 26.54,p = .047$, with a medium effect size, $\phi_c=.24$). 

Finally, it can be observed that LVHA emotion is associated with sharing less precise location information with Friends on Social Media among male and female participants with $\chi^2 (16, N = 60) = 27.30, p = .038$ and $\chi^2 (12, N = 60) = 22.50, p = .032$ respectively.

\begin{figure}[ht!]
    \centering
    \begin{subfigure}[b]{0.42\linewidth}
         \centering
        \includegraphics[width=\linewidth]{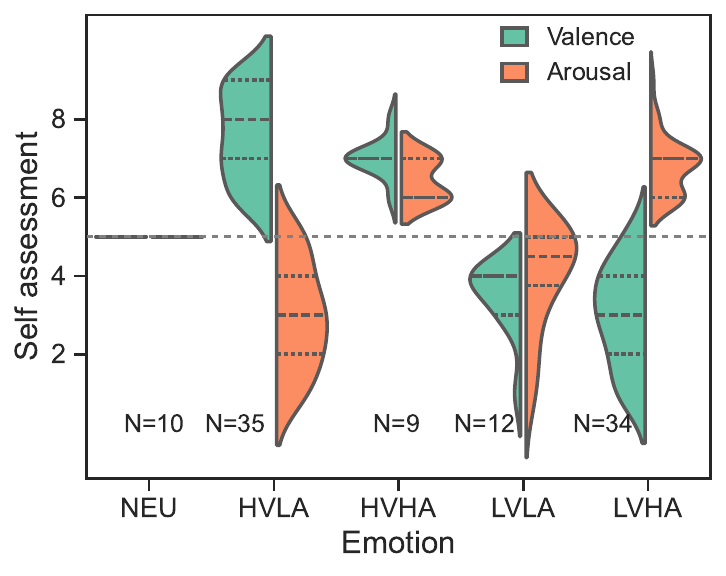}
     \end{subfigure}
     \begin{subfigure}[b]{0.52\linewidth}
          \centering
        \includegraphics[width=\linewidth]{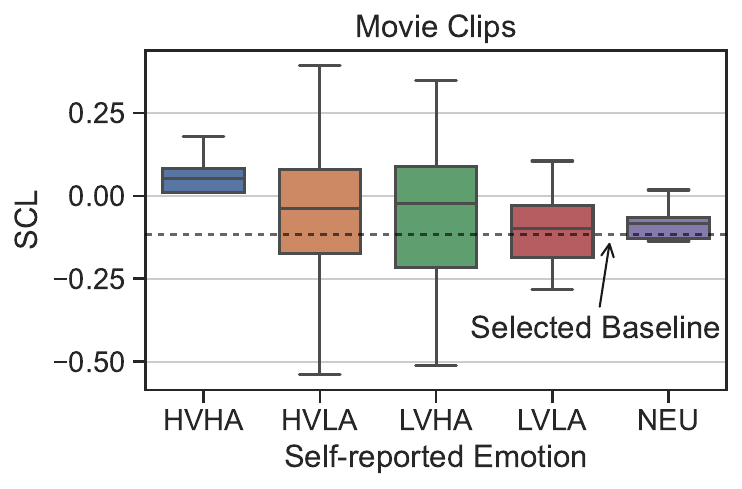}
     \end{subfigure}
    \caption{The distribution of the subjective ratings and Tonic EDA values (SCL) on the movie clips.}
    \label{fig:results_emotions}
\end{figure}

\paragraph{\textbf{Analysis of Physiological Variations}}
Similar to~\cite{hernelectrodermal2017}, EDA data collected from the E4 band is later mapped based on the SAM answers after watching the movie clips. As can be seen in Figure~\ref{fig:results_emotions}, the self-rated average at 7.0 valence and 6.5 arousal levels for `HVHA', 7.6 valence and 2.9 arousal for `HVLA',
3.0 valence and 6.8 arousal levels for `LVHA', and 
3.5 valence and 4.0 arousal levels for `LVLA'. Given varied perceptions of movie clips, for each participant, we selected the data collected that self-rated as the closest to neutral emotion (i.e., valence and arousal equal 5) as the baseline. The median SCL of the baseline is overall lower than the median of any self-reported emotions for the movie clips. In addition, `HVHA' has the highest values while `LVLA' has the lowest. 

\begin{figure}[ht!]
    \centering
    \begin{subfigure}[b]{0.47\linewidth}
         \centering
        \includegraphics[width=\linewidth]{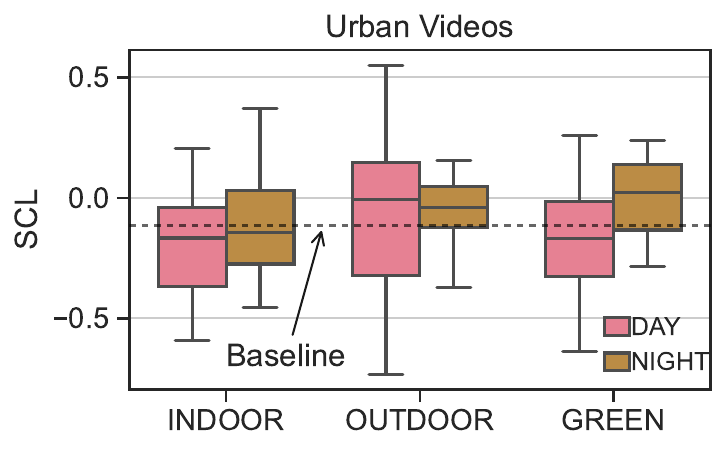}
     \end{subfigure}
     \hfill
     \begin{subfigure}[b]{0.47\linewidth}
          \centering
        \includegraphics[width=\linewidth]{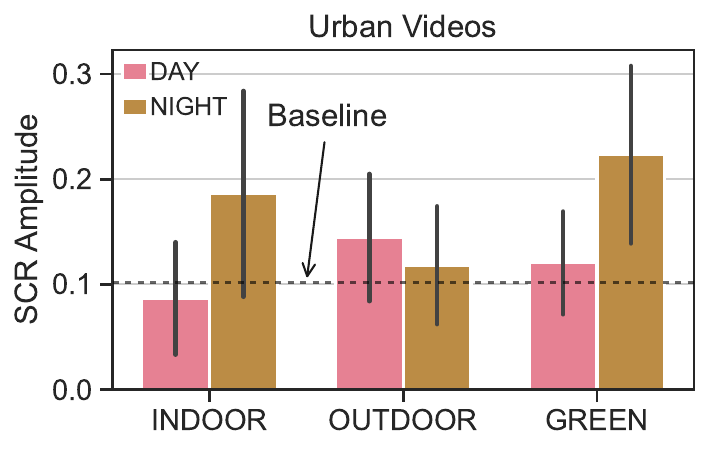}
     \end{subfigure}
    \caption{The distribution of the SCL and SCR amplitude on urban clips. Outliers have been removed for presentation purposes.}
    \label{fig:results_urban_emotion}
\end{figure}

When comparing these to the responses in urban clips (Figure~\ref{fig:results_urban_emotion}), we can see that when exposed to different urban environments at day and night time, the SCL values overall close to the baseline, with the SCR amplitudes much lower. This suggests mostly neutral and calm emotions, which is consistent with the self-ratings in Figure~\ref{fig:loc_emotion}. In the night clips, SCL and SCR amplitude reached the highest in greenspace, suggesting mostly negative and agitated emotions which is also aligned with self-ratings. 
\section{Discussion and Future Directions}

This study investigates the role of contextual emotions on personal location information sharing. Our observations suggests that 
distinct emotions have a more pronounced impact among social groups that are largely influenced by participants' GIPC.
Participants tend to share less precise location information with Close Friends and Family when feeling neutral, implying a cautious approach to privacy in familiar circles while their lower arousal levels lead to more precise location sharing with the University Community. Whilst these two social groups have been identified to be correlated with participants' general privacy attitudes, we have made an interesting observation regarding the Friends on Social Media category where high arousal lead to less precise sharing with this group, suggesting a protective attitude in broader, less intimate networks. These insights highlights the complex interplay between emotions and personal information sharing behaviours, emphasising the need for emotion-aware privacy settings. Moreover, our preliminary analysis of the  SCL and SCR amplitudes supports the idea that sensor data can provide continuous, objective measures of emotional responses to various contexts, allowing for more nuanced studies on how context and emotion influence personal information sharing behaviour. Understanding these dynamics informs creating future personalised strategies that can mitigate the risks associated with oversharing and can enhance user experience by tailoring content that respects users' emotional and contextual nuances.

\newpage
\bibliographystyle{ACM-Reference-Format}
\bibliography{reference}

\end{document}